\documentstyle[12pt]{article}
\newcommand{\eq}{\begin{equation}}
\newcommand{\en}{\end{equation}}
\newcommand{\eqn}{\begin{eqnarray}}
\newcommand{\enn}{\end{eqnarray}}
\newcommand{\CR}{\nonumber \\}
\newcommand{\A}{\alpha}
\newcommand{\B}{\beta}
\newcommand{\D}{\delta}
\newcommand{\DE}{\Delta}
\newcommand{\G}{\gamma}
\newcommand{\bg}{{\bf g}}
\newcommand{\hbg}{\hat{\bf g}}
\newcommand{\hbn}{\hat{\bf n}}
\newcommand{\hbh}{\hat{\bf h}}
\newcommand{\hI}{\hat{I}}
\newcommand{\hJ}{\hat{J}}
\newcommand{\hH}{\hat{H}}
\newcommand{\cM}{{\cal M}}

\newcommand{\cR}{{\cal R}}
\newcommand{\cW}{{\cal W}}

\newcommand{\tW}{\widetilde{W}}
\newcommand{\pa}{\partial}
\newcommand{\vp}{\varphi}
\begin{document}
\renewcommand{\thefootnote}{\fnsymbol{footnote}}
\begin{titlepage}
\null
\begin{flushright}
UTHEP-297 \\
March, 1995
\end{flushright}
\vspace{2.5cm}
\begin{center}
{\Large \bf
Free Field Realization of $WBC_{n}$ and $WG_{2}$ algebras
\par}
\lineskip .75em
\vskip 3em
\normalsize
{\large Katsushi Ito}
\vskip 1.5em
and
\vskip 1.5em
{\large Seiji Terashima}
\vskip 1.5em
{\it Institute of Physics, University of Tsukuba, Ibaraki 305, Japan}
\vskip 1.5em
\vskip 1.5em
{\bf Abstract}
\end{center} \par
We study the BRST-cohomology in the quantum hamiltonian reduction
of  affine Lie algebras of non-simply laced type.
We obtain the free field realization of
the $W{\bf g}$-algebra for $\bg=B_{2}$, $B_{3}$, $C_{3}$ and $G_{2}$.
The $WC_{3}$ algebra is shown to be equal to the $WB_{3}$
algebra at the quantum level by duality transformation.

\end{titlepage}
\renewcommand{\thefootnote}{\arabic{footnote}}
\setcounter{footnote}{0}
\baselineskip=0.7cm
$W$-algebra symmetry \cite{Za,BoSc} plays an important role in the
classification of rational conformal field theory and integrable systems
such as Toda field theory and 2d gravity.
The classical $W\bg$-algebra associated with a simple Lie algebra $\bg$ can be
realized as the second hamiltonian structure of the generalized KdV hierarchy
\cite{Walg}.
The classical hamiltonian reduction of affine Lie algebra provides a
systematic method to the construction of  generalized KdV hierarchy
associated with any Lie algebra\cite{DrSo}.
In order to study the quantum $W$ algebra
it is necessary to calculate operator product expansions of
normal-ordered composite operators.
The most general form of the $W$-algebra can be determined by
consistency conditions such as the Jacobi identity.
Such an approach has been done systematically for the $W$-algebras with
two and three generators \cite{KaWa,BFKNRV}.
However, it is technically difficult to extend this approach to
general $W$-algebra because of the lack of Lie algebraic viewpoint.

The free field realization of the $W\bg$ algebra is a
crucial step to understand the representation of the algebra,
correlation functions through screening operators, which is defined
as the  commutant  of the $W$-algebra.
So far the free field realization is well-understood for the $W\bg$-algebra
for simply laced Lie algebras $\bg=A_{n}$ and $D_{n}$\cite{FaLu}.
The $W\bg$-algebra based on a simply laced Lie algebra
has self-dual property in the torus, on which
free bosons are compactified.
Self-duality allows us to generalize the classical Miura transformation
to the quantum one by taking normal ordered product of the scalar Lax
operator and replacing the level $k$ of an affine Lie
algebra by a parameter $\A_{0}=a-1/a$ ,
where $a=\sqrt{k+h^{\vee}}$ and $h^{\vee}$ is the dual Coxeter number.

Concerning the quantum $W\bg$ algebra associated with
a non-simply laced Lie algebra $\bg$, it gets non-trivial quantum correction
due to the lack of self-duality.
In fact, the classical Miura transformation for non-simply laced Lie
algebra does not work in the quantum case.
The purpose of the present article is to study the free field
realization of the quantum $W$-algebra associated with non-simply
laced Lie algebras.
We will use the method of quantum hamiltonian reduction
\cite{BeOo,FeFr,BoTj} based on the BRST quantization.
We examine the BRST cohomology explicitly for
non-simply laced Lie algebras $B_{2}$, $B_{3}$, $C_{3}$ and $G_{2}$.
An interesting properties in non-simply laced Lie algebra is
the duality relation between Lie algebras $B_{n}$ and $C_{n}$.
This duality predicts a unique quantum $W$ algebra for $BC$ type
Lie algebra\cite{FeFr2}.
We will confirm this duality for $BC_{3}$ case.

Let $\bg$ be a simple Lie algebra $\bg$ with rank $r$.
$\DE$ the set of roots $\bg$, $\DE_{+}$ the set of positive
roots and $\A_{1}, \ldots, \A_{r}$ the simple roots.
Let $\{ t_{a} \}$ be a basis of $\bg$, satisfying
commutation relations $\mbox{[} t_{a}, t_{b} \mbox{]}=f_{a,b}^{c}t_{c}$,
where $f_{a,b}^{c}$ are the structure constants.
In the following we shall use the Chevalley canonical basis:
$\{ t_{a} \}=\{ e_{\A}, h_{i} \}$ $(\A\in \DE, i=1,\ldots, r)$, with
commutation relations:
\eqn
\mbox{[} e_{\A}, e_{\B} \mbox{]} &=& \left\{ \begin{array}{ll}
 N_{\A,\B} e_{\A+\B}, &
\mbox{for $\A+\B\in \DE$} \\
{2\A\cdot h \over \A^{2}},& \mbox{for $\A+\B=0$}
\end{array} \right. \CR
\mbox{[} h_{i}, e_{\A} \mbox{]} &=& \A^{i} e_{\A}
\enn
The metric $g_{a,b}$ is defined by
$g_{\A,\B}=2\D_{\A+\B,0}/\A^{2}$ and $g_{i,j}=\D_{i,j}$,
which are used to raise and lower indices of the generators.
Let $\{ x_{0}, e_{0}, f_{0} \}$ ($\mbox{[}e_{0},f_{0}\mbox{]}=x_{0},
\mbox{[}x_{0},e_{0} \mbox{]}=e_{0}, \mbox{[}x_{0},f_{0} \mbox{]}=-f_{0}$)
be a $sl(2)$ subalgebra of $\bg$.
The principal embedding of $sl(2)$ into $\bg$ is characterized by
$e_{0}=\sum_{i=1}^{r}e_{\A_{i}}$\cite{Ko}.
With respect to the principal $sl(2)$ embedding $\bg$ may be
decomposed as $\bg=\oplus_{k} \bg_{k}$, where $\bg_{k}\equiv
\{ x\in \bg; ({\rm ad} x_{0}) x=k x\}$
is a $2k+1$ dimensional subspace in $\bg$ and $k$
runs over the exponents of $\bg$.

The affine Lie algebra $\hbg$ at level $k$ is generated by the
currents $\{ J_{a}(z) \}\!=\!\{ J_{\A}(z), H_{i}(z)\} $ $(\A\in \DE,
i=1,\ldots, r$) satisfying the operator product expansions:
\eq
J_{a}(z)J_{b}(w)={k g_{ab} \over (z-w)^{2}}
                   +{ f_{a b}^{c} J_{c}(w) \over z-w}+\cdots .
\en
$\hbg$ admits a triangular decomposition
$\hbg=\hbn_{+}\oplus\hbh\oplus\hbn_{-}$, where
the algebra $\hbn_{+}$ ( $\hbn_{-}$ ) is  generated by the
currents which correspond to the positive (negative) roots of $\bg$,
$\hbh$ by the Cartan currents $H_{i}(z)$.

Let us consider the constraints for the currents associated with
the principal $sl(2)$ embedding into $\bg$:
\eq
\chi(J^{\A}(z))= \left\{ \begin{array}{ll}
                    1 & \mbox{for $\A=\A_{i}$,\ \  $i=1, \ldots, r$} \\
                    0 & \mbox{for $\A\in\DE_{+}$: non-simple roots.}
                    \end{array} \right.
\en
Denote $\cM$ the phase space with the above constraint.
We may consider the reduced phase space $\cR$ by taking quotient space
with respect to the residual gauge symmetry generated by $\hbn_{+}$.
Classically, $\cR$ carries the Poisson bracket structure.
By choosing a special gauge, we have the representation
of the classical $W$-algebra\cite{DrSo}.

We shall use the BRST formalism to quantize this system.
We introduce fermionic ghost fields
$(b^{\A}(z),c_{\A}(z))$ ($\A\in\DE_{+}$) with operator product expansion:
$
b^{\A}(z)c_{\A'}(w)=\D^{\A}_{\A'}/(z-w)+\cdots.
$
Define the BRST charge:
\eq
Q_{BRST}=\int { dz \over 2\pi i} J_{BRST}(z),
\en
where $J_{BRST}(z)$ is the BRST current
\eq
J_{BRST}(z)=\sum_{\A\in\DE_{+}}c_{\A}(J^{\A}-\chi(J^{\A}))(z)
           -{1\over 2}\sum_{\A,\B,\G\in\DE_{+}}f^{\A,\B}_{\G}
            (c_{\A}(c_{\B}b^{\G}))(z).
\en
The normal ordered product $(AB)(z)$ for operators
$A(z)$ and $B(z)$ is defined by
\eq
(AB)(z)=\int_{z} {d w\over 2 \pi i} {A(w)B(z)\over w-z}.
\en
The BRST operator $Q_{BRST}$ satisfies the nilpotency condition
$Q_{BRST}^{2}=0$.
The physical Hilbert space may be characterized by investigating the
$Q_{BRST}$-cohomology.

The information of the $W$-algebra is contained
in the cohomology on the space of the
universal enveloping algebra $U(\hbg)$ and the Clifford algebra
$Cl_{b,c}$
generated by the fermionic ghost fields.
We decompose the BRST current into two parts:
$J_{BRST}(z)=J_{0}(z)+J_{1}(z)$, where
\eqn
J_{0}(z)&=&\sum_{\A\in\DE_{+}}c_{\A}J^{\A}(z)
           -{1\over 2}\sum_{\A,\B,\G\in\DE_{+}}f^{\A,\B}_{\G}
            (c_{\A}(c_{\B}b^{\G}))(z), \CR
J_{1}(z)&=&-\sum_{\A\in\DE_{+}}c_{\A}\chi(J^{\A})(z).
\enn
Define fermionic charges $Q_{i}$ by the contour integration of
$J_{i}(z)$.
Since these charges obey the relations
$(Q_{0})^{2}=(Q_{1})^{2}=\{ Q_{1}, Q_{0} \}=0$, we
may use the spectral sequence technique for the double complex
generated by $Q_{0}$ and $Q_{1}$\cite{BoTu}.
The operator $Q_{0}$ is the canonical coboundary operator of the
nilpotent subalgebra $\hbn_{+}$.
The operator $Q_{1}$ defines a gradation in the BRST complex
associated with the principal embedding $sl(2)$ into $\hbg$.
Feigin and Frenkel analyzed this complex by taking $Q_{0}$-cohomology
first\cite{FeFr}.
Under the $Q_{0}$-cohomology, the $Q_{1}$-cohomology reduces to the
problem of finding conserved currents in quantum Toda field theory
\cite{KaWa2}.
On the other hand, De Boer and Tjin studied the BRST-complex by
taking the $Q_{1}$-cohomology first\cite{BoTj}.
They observed that the components of non-trivial cohomology which
have zero gradation in total degree of the double complex form
a closed algebra and the generators are nothing but a free field
realization.

For the analysis of the BRST cohomology, it is convenient to introduce
new currents
\eq
\hat{J}^{a}(z)=J^{a}(z)+f^{a \B}_{\G} (b^{\G}c_{\B})(z)
\en
modified by ghost fields.
We decompose the set of modified currents $\{ \hat{J}^{a} \}$ into
the submodules $\oplus_{k}V_{k}$ under the principal $sl(2)$
embedding, where $k$ belongs to the set of exponents of $\bg$.
$V_{k}$ denotes the set of currents
$\{ \hat{I}^{k}_{m} \}_{m=-k, -k+1, \ldots,k-1,k}$.

{}Since $Q_{BRST}(b^{\A})=\hat{J}^{\A}-\chi(J^{\A})$
and $ b^{\A} $ is a non-trivial element in the BRST complex,
the pairs $(b^{\A}, \hat{J}^{\A}-\chi(J^{\A}))$ form the BRST doublets
and decouples from the non-trivial cohomology.
Therefore we consider the BRST-cohomology on the
reduced complex ${\cal A}_{red}$, which is
spanned by other modified currents and ghost fields.
First we study the $Q_{1}$-cohomology $H_{Q_{1}}({\cal A}_{red})$
on the reduced complex ${\cal A}_{red}$.
Since the lowest component $\hat{I}^{k}_{-k}$ belongs to the
kernel of $Q_{1}$, we can choose $\hat{I}^{k}_{-k}$ as the basis of
$H_{Q_{1}}({\cal A}_{red})$.
We put $\cW^{(0)}_{k+1}=\hat{I}^{k}_{-k}$.
In order to obtain observables of the full BRST-cohomology,
we need to solve the descent equation:
\eq
Q_{0}({\cal W}^{(n)}_{k+1})=Q_{1}({\cal W}^{(n+1)}_{k+1}).
\label{eq:des}
\en
The generators of the total BRST cohomology are given by
\eq
{\cal W}_{k+1}={\cal W}^{(0)}_{k+1}-{\cal W}^{(1)}_{k+1}+{\cal W}^{(2)}_{k+1}
+\cdots+(-1)^{k} {\cal W}^{(k)}_{k+1} ,
\en
where $Q_{0}({\cal W}^{(k)}_{k+1})=0$.

Now we consider the BRST complex for $B_{n}^{(1)}$ case.
The positive roots system of Lie algebra $B_{n}$ is
$e_{i}-e_{j}$, $e_{i}+e_{j}$ ($i<j$) and $e_{i}$, where we
introduced an orthonormal basis  $e_{i}$ ($i=1,\ldots, n$) satisfying
$e_{i}\cdot e_{j}=\D_{i,j}$.
The fundamental representation of $B_{n}$ is given by $(2n+1)\times (2n+1)$
matrix of the form
\eqn
e_{e_{i}-e_{j}}&=& E_{i,j}-E_{2n+2-j, 2n+2-i}, \CR
e_{e_{i}+e_{j}}&=& E_{i,2n+2-j}-E_{j,2n+2-i}, \CR
e_{e_{i}}&=& \sqrt{2}(E_{i,n+1}-E_{n+1,2n+2-i})
\enn
and $e_{-\A}={}^{t}e_{\A}$,
$2\A\cdot h/\A^{2}=\mbox{[}e_{\A},e_{-\A}\mbox{]}$.
The structure constants can be easily calculated from this representation.

We discuss the $B_{2}$ case for simplicity.
The principal $sl(2)$ decomposition $V_{1}\oplus V_{3}$ of $B_{2}^{(1)}$
is given by
\eqn
\hI^{1}_{1}&=& \hJ^{e_{2}}+\hJ^{e_{1}-e_{2}}, \quad
\hI^{1}_{0}=(2 e_{1}+e_{2})\cdot\hH, \quad
\hI^{1}_{-1}=3\hJ^{-e_{2}}+4\hJ^{-(e_{1}-e_{2})} \CR
\hI^{3}_{\pm 3}&=& \hJ^{\pm(e_{1}+e_{2})},\quad
\hI^{3}_{\pm 2}=  \hJ^{\pm e_{1}}, \CR
\hI^{3}_{1}&=&2\hJ^{e_{2}}-3\hJ^{e_{1}-e_{2}}, \quad
\hI^{3}_{-1}=\hJ^{-e_{2}}-2\hJ^{-(e_{1}-e_{2})}, \quad
\hI^{3}_{0}=(e_{1}-2 e_{2})\cdot \hH .
\enn
The $Q_{1}$-cohomology is generated by
$\cW_{2}^{(0)}= \hI^{1}_{-1}$ and $\cW_{4}^{(0)}= \hI^{3}_{-3}$.
By solving the descent equation (\ref{eq:des}), we get
\eqn
{\cal W}_{2}&=& \cW_{2}^{(0)}-\cW_{2}^{(1)}, \CR
\cW_{4}&=& \cW_{4}^{(0)}-\cW_{4}^{(1)}+\cW_{4}^{(2)}-\cW_{4}^{(3)},
\enn
where
\eqn
\cW_{2}^{(1)}&=& {1\over 20}
                 \left( (23+10k)\pa\hI^{1}_{0}+\pa\hI^{3}_{0}
                        +5(\hI^{1}_{0}\hI^{1}_{0})+5(\hI^{3}_{0}\hI^{3}_{0})
                 \right), \CR
\cW_{4}^{(1)}&=& -({11\over 5}+k) \pa \hI^{3}_{-2}
                 +3(\hI^{1}_{0}\hI^{3}_{-2})-(\hI^{3}_{0}\hI^{3}_{-2})
                 +(\hI^{3}_{-1}\hI^{3}_{-1})
                 -6 (\hI^{3}_{-1}\hI^{1}_{-1}), \CR
\cW_{4}^{(2)}&=& (-{291\over 50}-5k-k^{2}) \pa^{2}\hI^{3}_{-1} \CR
             & &     +{18\over 5} (\hI^{1}_{0}\pa\hI^{1}_{-1})
                 -(14+5k)(\hI^{3}_{0}\pa\hI^{1}_{-1})
                 +(11+5k) (\hI^{1}_{0}\pa\hI^{3}_{-1})
                 -{2\over 5} (\hI^{3}_{0}\pa\hI^{3}_{-1}) \CR
             & & -{24\over 5} (\hI^{1}_{-1}\pa\hI^{1}_{0})
                 -({83\over 5}+4k) (\hI^{1}_{-1}\pa\hI^{3}_{0})
                 +(5+2k) (\hI^{3}_{-1}\pa\hI^{1}_{0})
                 +({17\over 5}+k) (\hI^{3}_{-1}\pa\hI^{3}_{0})\CR
             & & +18 (\hI^{1}_{0}(\hI^{3}_{0}\hI^{1}_{-1}))
                 -(\hI^{3}_{0}(\hI^{3}_{0}\hI^{1}_{-1}))
                 -6 (\hI^{1}_{0}(\hI^{1}_{0}\hI^{3}_{-1}))
                 -(\hI^{1}_{0}(\hI^{3}_{0}\hI^{3}_{-1}))
                 +(\hI^{3}_{0}(\hI^{3}_{0}\hI^{3}_{-1})), \CR
\cW_{4}^{(3)}&=& (63+85k +25 k^{2}) \pa^{3}\hI^{1}_{0}
                -({714\over 125}+{917k\over 120}
         +{407k^{2}\over 120}+{k^{3}\over 2}) \pa^{3}\hI^{3}_{0} \CR
             & & +{3(13+5k)\over 50} (\hI^{1}_{0}\pa^{2}\hI^{1}_{0})
                 +{3(20+18k+4 k^2)\over4} (\hI^{1}_{0}\pa^{2}\hI^{3}_{0})
             +{20+18k+4 k^2\over 8} (\hI^{3}_{0}\pa^{2}\hI^{1}_{0})\CR
             & &    + ({82\over 25}+{51 k\over 20}+{k^2\over 2})
                   (\hI^{3}_{0}\pa^{2}\hI^{3}_{0})
                   +({1091\over100}+{46 k\over5}+2 k^2)
                    (\pa\hI^{1}_{0}\pa\hI^{3}_{0})\CR
             & & +({281\over 50}+{415 k\over100} +{150 k^2\over200})
                  (\pa\hI^{3}_{0}\pa\hI^{3}_{0})
             -({69\over25}+{6 k\over5}) (\pa\hI^{1}_{0}\pa\hI^{1}_{0}) \CR
             & &  -(9+3k) (\hI^{1}_{0}(\hI^{3}_{0}\pa\hI^{3}_{0}))+
                 ({69\over10}+3k) (\hI^{3}_{0}(\hI^{3}_{0}\pa\hI^{3}_{0}))
              + {3\over 10} (\hI^{1}_{0}(\hI^{1}_{0}\pa\hI^{1}_{0})) \CR
         & &    -(8+{7k\over 2}) (\hI^{1}_{0}(\hI^{3}_{0}\pa\hI^{1}_{0}))+
          ({1\over 20}-{k\over2}) (\hI^{3}_{0}(\hI^{3}_{0}\pa\hI^{1}_{0}))
         -{247+110 k\over 20} (\hI^{1}_{0}(\hI^{1}_{0}\pa\hI^{3}_{0}))\CR
          & & + 3 (\hI^{1}_{0}(\hI^{1}_{0}(\hI^{1}_{0}\hI^{3}_{0})))+
                {7\over 4} (\hI^{1}_{0}(\hI^{1}_{0}(\hI^{3}_{0}\hI^{3}_{0})))
                -3 (\hI^{1}_{0}(\hI^{3}_{0}(\hI^{3}_{0}\hI^{3}_{0}))).
\enn
The last components $\cW_{2}^{(2)}$ and $\cW_{4}^{(3)}$
in the double complex are expressed in terms of zero degree fields
$\hI^{1}_{0}$ and  $\hI^{3}_{0}$.
Introduce free bosons $\vp_{i}(z)$ ($i=1,2$) by
$i a \pa \vp_{j}=e_{j}\cdot \hat{H}$ where $a=\sqrt{k+3}$ and
$\vp_{i}(z)\vp_{j}(w)=-\log (z-w) +\cdots$.
One finds that the fields
\eqn
T(z)&=& {40 \over a^{2}} \cW_{2}^{(2)}(z), \CR
W_{4}(z)&=& -{25\over a^{4}} \cW_{4}^{(3)}(z)
\enn
generates the $WB_{2}$ algebra \cite{HaTa} with
the central charge
\eq
c=86-{30\over a^2}-60 a^2={-2(12+5k) (13+6k)\over 3+k}.
\en
In terms of free fields, the generators of the $WB_{2}$ algebra are
expressed as
\eqn
T(z)&=& {1\over2}(p_{1}^2+p_{2}^2)
       +({3\over 2 a}-2 a)\pa p_{1}+({1\over 2a}-a)\pa p_{2}, \CR
\!\!\!W_{4}(z)&=& -p_{1}^4 -p_{2}^4
        +{17\over4}p_{1}^2 p_{2}^2
      +2(-{3\over a}+4a) p_{1}^2\pa p_{1}
        +17({1\over 4 a}-{a\over 2}) p_{1}^2\pa p_{2} \CR
    & & +25({1\over 2 a}-{a\over 2}) p_{1}p_{2}\pa p_{2}
        +(-{2\over a}+4 a)p_{2}^2\pa p_{2}
        +({1\over 4a}-{9a\over 2}) (\pa p_{1})p_{2}^2\CR
    & & +({39\over4}-{63\over 20 a^2}-{15 a^2\over 2})p_{1}\pa^2 p_{1}
        +(-{75\over4}+{25\over4a^2}+{25a^2\over2})p_{1}\pa^2 p_{2} \CR
    & & +(-9 +{31\over 10 a^2}+5a^2)p_{2}\pa^2 p_{2}
        +({43\over 4}-{59\over 10 a^2}-{19a^2\over 4})(\pa p_{1})^2 \CR
    & & +(-5 +{1\over 4a^2}+9a^2)(\pa p_{1})(\pa p_{2})
        +(-3+{21\over 10 a^2}+a^2) (\pa p_{2})^2 \CR
  \!\!\!& &\!\!\! +(-{8\over 5a^3}+{631\over 120a}
      -{145a\over24}+{5a^3\over2})\pa^3 p_{1}
      +({31\over 20 a^3}-{109\over 15 a}+{65a \over 6}-5a^3 )\pa^3 p_{2},
\enn
where $p_{i}=i\pa\vp_{i}$.
Note that $W_{4}(z)$ is a quasi-primary field.
Spin four primary field $\widetilde{W}_{4}(z)$ is defined by
\eq
\widetilde{W}_{4}=W_{4}+{{7 \left( 75 - 113 {a^2} \right) }\over
   {4 \left( 75 - 226 {a^2} + 150 {a^4} \right) }}(TT)
   +{{1950 - 9101 {a^2} + 11245 {a^4} - 3300 {a^6}}\over
   {40 {a^2} \left( 75 - 226 {a^2} + 150 {a^4} \right) }}\pa^2 T.
\en
The present free field realization agrees with that obtained in ref.
\cite{KaWa2}.

By a similar procedure we may construct the free field realization of
$WB_{3}$-algebra.
We define the energy-momentum tensor
$T(z)$ and quasi-primary fields $W_{4}(z)$ and $W_{6}(z)$
with spins 4 and 6, respectively as follows:
\eqn
T(z)&=& -{28\over a^{2}}\cW_{2}^{(1)}(z), \CR
W_{4}(z)&=& -{48\over a^{4}} \cW_{4}^{(3)}(z), \CR
W_{6}(z)&=& {32928 \over 235 a^{6}} \cW_{6}^{(5)}(z).
\enn
where  $a=\sqrt{k+5}$.
The central charge is $c=267-105/a^{2}-168 a^{2}$.
As in the case of $WB_{2}$ ,  we introduce free bosons
$p_{i}\equiv e_{i}\cdot\hat{H}/a$.
The free field realization of $T(z)$ and $W_{4}$ is given by
\eqn
T(z)&=&
{{{{p_{1}}^2}}\over 2} + {{{{p_{2}}^2}}\over 2} + {{{{p_{3}}^2}}\over 2} +
  ({5\over 2a}-3a)\pa p_{1}
  +({3\over2a}-2a)\pa p_{2}
 +({1\over 2a}-a) \pa p_{3}, \CR
W_{4}(z)&=&{{p_{1}}^4} - 2 {{p_{1}}^2} {{p_{2}}^2} + {{p_{2}}^4} -
  2 {{p_{1}}^2} {{p_{3}}^2} - 2 {{p_{2}}^2} {{p_{3}}^2} + {{p_{3}}^4}\CR
& &  + 2({5\over a}-6a) {{p_{1}}^2} (\pa p_{1})
     - 2({1\over a}-2a) {{p_{2}}^2} (\pa p_{1})
     -2({1\over a}-2a) {{p_{3}}^2} (\pa p_{1}) \CR
& &
     -2({3\over a} -4a){{p_{1}}^2} (\pa p_{2})
     -8({1\over a}-a) p_{1} p_{2} (\pa p_{2})
     +2({3\over a}-4a) {{p_{2}}^2} (\pa p_{2})
     +{2\over a} {{p_{3}}^2} (\pa p_{2}) \CR
& & -2({1\over a}-2a) {{p_{1}}^2} (\pa p_{3})
    -2({1\over a}-2a) {{p_{2}}^2} (\pa p_{3})
    -8({1\over a}-a) p_{1} p_{3} (\pa p_{3}) \CR
& &     -8({1\over a}-a)p_{2} p_{3} (\pa p_{3})
    +2({1\over a}-2a) {{p_{3}}^2} (\pa p_{3})
+{(5 + 2 a^2 ) (17 - 42 a^2 )\over 7 a^{2}}
      {{(\pa p_{1})}^2} \CR
& &      +{2(3-4a^{2})(-1+2a^{2})\over a^{2}} (\pa p_{1}) (\pa p_{2})
    -({64\over 7}-{29\over 7 a^{2}}-4 a^{2}) {(\pa p_{2})}^2 \CR
& & +(8-{2\over a^{2}}-8 a^{2}) (\pa p_{1}) (\pa p_{3})
    -(4 -{2\over a^{2}}) (\pa p_{2}) (\pa p_{3})
    +({48\over 7}-{27\over 7 a^{2}} -4 a^{2}) {(\pa p_{3})}^2 \CR
& & +(-{190\over 7}+{78\over 7 a^{2}}+16 a^{2}) p_{1} (\pa^{2}p_{1})
    +(28-{12\over a^{2}}-16 a^{2}) p_{1} \pa^{2}p_{2}
    +({6\over 7}-{6\over 7 a^{2}}) p_{2} \pa^{2}p_{2} \CR
& &    +(12 -{4\over a^{2}}-8 a^{2}) p_{1} \pa^{2}p_{3}
    +(12 -{4\over a^{2}}-8 a^{2}) p_{2} \pa^{2}p_{3}
    +({90\over 7}-{34\over 7 a^{2}}-8 a^{2}) p_{3} \pa^{2}p_{3} \CR
& &
 +({55\over 7 a^{3}}-{538\over 21 a}+{548 a \over 21}-8 a^{3}) \pa^{3}p_{1}
 +(-{23\over 7 a^{3}}+{38\over 3 a}-{358 a \over 21}+8 a^{3})
  \pa^{3}p_{2} \CR
& &
 +(-{17\over 7 a^{3}}+{230 \over 21 a}-{340 a \over 21}+8 a^{3})
  \pa^{3}p_{3}.
\enn
Note that the $W$-current $W_{\DE}(z)$ of spin $\DE$ can be expressed as
\eq
W_{\DE}(z)=
\sum_{i_{1}, \cdots }
d^{\DE}(1^{i_{1}}2^{i_{2}}\cdots, 1^{j_{1}}2^{j_{2}}\cdots)
p_{1}^{i_{1}}(\pa p_{1})^{i_{2}} \cdots
           p_{2}^{j_{1}}(\pa p_{2})^{j_{2}} \cdots
\en
in terms of free fields $p_{j}=i\pa\vp_{j}$ ($j=1, \ldots $).
Here $i_{1}, i_{2} \ldots$ are non-negative integers and satisfy
\eq
\DE=i_{1}+2 i_{2} +\cdots +j_{1}+2 j_{2}+\cdots.
\en
The list of non-zero coefficients of the spin 6 field $W_{6}(z)$
is given in Table 1.

The spin 4 and 6 primary $W$-currents $\tW_{4}(z)$ and $\tW_{6}(z)$
are defined by
\eqn
\widetilde{W}_{4}&=&W_{4}-
{4 \left( 81 - 113 {a^2} \right) \over 525 - 1357 {a^2} + 840 {a^4}}
(TT)
   -{2 \left( 2835 - 11973 {a^2} + 16138 {a^4} - 6888 {a^6} \right) \over
   {7 {a^2} \left( 525 - 1357 {a^2} + 840 {a^4} \right)} }
\pa^2 T, \CR
\widetilde{W}_{6}&=&W_{6}+b_{1}(\tW_{4}T)+b_{2}(T(TT))
+b_{3}(T\pa^{2}T)+b_{4}(\pa T\pa T)+b_{5} \pa^{4}T +b_{6} \pa^{2}\tW_{4},
\enn
where
{\small
\eqn
b_{1}\!\!\!&=&\!\!\! {{686\,\left( -11 + 17\,{a^2} \right) }\over
        {141\,\left( 35 - 97\,{a^2} + 56\,{a^4} \right) }} , \CR
b_{2}\!\!\!&=&\!\!\!{{24\,\left( -53482275 + 302341760\,{a^2} -
             644454069\,{a^4} + 615107512\,{a^6} - 222277440\,{a^8} \right) }
          \over {235\,\left( -35 + 48\,{a^2} \right) \,
           \left( 525 - 1357\,{a^2} + 840\,{a^4} \right) \,
           \left( 735 - 1937\,{a^2} + 1176\,{a^4} \right) }}, \CR
b_{3}\!\!\!&=&\!\!\!{ 2\over
        {1645\,{a^2}\,\left( -35 + 48\,{a^2} \right) \,
          \left( 525 - 1357\,{a^2} + 840\,{a^4} \right) \,
          \left( 735 - 1937\,{a^2} + 1176\,{a^4} \right) }} \CR
 & & \!\!\!\!\!\!\!\!\!\times
       {\left( 12915159075 - 62611596075\,{a^2} +
            82595313577\,{a^4} + 52890458027\,{a^6} - 221412298156\,{a^8}
       \right. }\CR
       & & {\left. +
            188135793216\,{a^{10}} - 52327860480\,{a^{12}} \right)}, \CR
b_{4}\!\!\!&=&\!\!\!{2\over
       {1645\,{a^2}\,\left( -35 + 48\,{a^2} \right) \,
         \left( 525 - 1357\,{a^2} + 840\,{a^4} \right) \,
         \left( 735 - 1937\,{a^2} + 1176\,{a^4} \right) }}\CR
& & \!\!\!\!\!\!\!\!\!
\times\left( 4981954950 - 38746440075\,{a^2} +
           118960035187\,{a^4} - 179703825989\,{a^6} + 133280698775\,{a^8}
\right. \CR
& & \left.  -
           38960324928\,{a^{10}} + 271656000\,{a^{12}} \right), \CR
b_{5}\!\!\!&=&\!\!\!{1\over
      {9870\,{a^4}\,\left( -35 + 48\,{a^2} \right) \,
        \left( 525 - 1357\,{a^2} + 840\,{a^4} \right) \,
        \left( 735 - 1937\,{a^2} + 1176\,{a^4} \right) }}
\Bigl( 162002673000 \CR
& &
  -\! 1435942360125\,{a^2} +\! 5197128910680\,{a^4} -\!
        9680172818260\,{a^6} +\! 9226997505872\,{a^8}  \CR
& & \left.-
        2865622277231\,{a^{10}} - 2298232755504\,{a^{12}} +
        2260116476352\,{a^{14}} - 566577607680\,{a^{16}} \right), \CR
b_{6}\!\!\!&=&\!\!\!
{{49\,\left( -2415 + 12328\,{a^2} - 16177\,{a^4} + 4704\,{a^6} \right)}
     \over {2115\,{a^2}\,\left( 35 - 97\,{a^2} + 56\,{a^4} \right) }} .
\enn
}
The $WB_{3}$ algebra is consistent with the third solution
of the $W_{4,6}$ algebra in Ref. \cite{KaWa},
which was calculated from the Jacobi identities.

We have also calculated the BRST-cohomology for an affine Lie algebra
$C_{3}^{(1)}$ .
A Lie algebra $C_{n}$ has simple roots
$\A_{i}=(e_{i}-e_{i+1})/\sqrt{2}$, ($i=1,\cdots, n-1$)
and $\A_{n}=\sqrt{2}e_{n}$, which are equal to simple co-roots of $B_{n}$.
The cohomological analysis shows that
the observables exist at spin 2,4 and 6.
Spin 4 and 6 quasi-primary fields $\cW_{4}^{(3)}$, $\cW_{6}^{(5)}$
are apparently different from those of $B_{3}^{(1)}$.
However, if one makes primary fields from these quasi-primary fields,
one finds that the $W$-currents are equal to those of $B_{3}^{(1)}$
by the duality transformation $a\rightarrow -\sqrt{2}/a$.

For an exceptional Lie algebra $G_{2}$, we may compute its structure
constant by considering the seven-dimensional representation,
which can be embedded into the vector representation of $B_{3}$:
\eqn
e_{\A_{1}}&=& E_{23}-E_{56},\CR
e_{\A_{2}}&=& E_{12}-E_{67}+\sqrt{2}(E_{34}-E_{45}).
\enn
Here
$
\A_{1}= e_{1}-e_{2}$ and
$
\A_{2}= e_{2}-{1\over3}(e_{1}+e_{2}+e_{3})
$
are simple roots.
By investigationg the BRST-cohomology,
one finds that generators of the $WG_{2}$ algebra are
\eqn
T(z)&=&{-28\over a^{2}}\cW_{2}^{(1)}(z), \CR
W_{6}(z)&=& \cW_{6}^{(5)}(z).
\enn
By introducing free fields
$\sqrt{2}p_{1}={\A_{1}\cdot \hat{H}/a}$ and
$\sqrt{3/2}p_{2}={\lambda_{2}\cdot \hat{H}/a}$, where $\lambda_{i}$ denotes
the fundamental weight of $G_{2}$ and $a=\sqrt{k+4}$,
the energy momentum tensor becomes
\eq
T={1\over2}(p_{1}^2+p_{2}^2)
       +{1\over \sqrt{2}}({1\over a}- a)\pa p_{1}
       +({5\over \sqrt{6}a}-3\sqrt{3\over2}a)\pa p_{2}.
\en
The spin 6 field $W_{6}$ is shown in Table 2.
We may construct primary field as
\eq
\widetilde{W}_{6}(z)=a_{1} W_{6}
+a_{2} (T(TT))
+a_{3} (T\pa^{2}T)
+a_{4} (\pa T\pa T)
+a_{5}  \pa^{4}T
\en
where
{\small
\eqn
a_{1}\!\!\!&=&\!\!\!
 (112 - 387 a^2  + 336 a^4 ) (196 - 713 a^2  + 588 a^4 ), \CR
a_{2}\!\!\!&=&\!\!\!
 32(-6278272 + 33345830 a^2  - 58292949 a^4  + 33559974 a^6)/ 27,
\CR
a_{3}\!\!\!&=&\!\!\!
 (-18461632 - 223338388 a^2  + 1362709749 a^4  - 1927045029 a^6  +
 34038144 a^8 \CR
& &   + 939880368 a^{10})/1323, \CR
a_{4}\!\!\!&=&\!\!\!
(156320192 - 1352237488 a^2  + 4934160444 a^4  - 9498512667 a^6
 +9458326773 a^8 \CR
 & &  - 3791786796 a^{10})/1323, \CR
a_{5}\!\!\!&=&\!\!\!
 (-1445670912 + 18242353472 a^2  - 98764008700 a^4
+ 294933524139 a^6,  \CR
& & -522573338406 a^8  + 549763546437 a^{10}   - 319548026700 a^{12}
 + 79796332224 a^{14})
  / 444528. \CR
\enn
}
The currents $(T,\widetilde{W}_{6})$ form a
closed algebra, which is
consistent with the previous Jacobi-identity analysis \cite{KaWa,BFKNRV}.

In the present article, we have examined the quantum hamiltonain
reduction of non-simply affine Lie algebra with rank two and three.
We have explicitly constructed higher spin currents of the $W$-algebra
in term of free bosons.
Altough it is still technically difficult to
generalize the present approach to arbitrary affine Lie algebra,
it would be possible to obtain the free field realization of the
$W$-currents with spin three or four.
In order to study the representation of the quantum $W$-alagbera, we need
to introduce screening operators.
The unitary representation of the $WBC_{n}$ algebra is particularly
interesting subject since it does not seem to correspond to any
coset construction \cite{It}.
We may also consider various generalization of the present
quantum hamiltonian reduction
( non-principal $sl(2)$
embeddings, affine Lie superalgebras etc. ) \cite{QHR}.
We have observed that the present quantum hamiltonian reduction
gives a natural generalization of the classical
Drinfeld-Sokolov approach, which can be applicable to any
affine Lie algebra.
Hence it is natural to ask how one can introduce an spectral
parameter in the quantum hamiltonian reduction,
which would give an explict and systematic method to construct the
qunatum conserved currents in a massive integrable field theory\cite{Za2}.
These subjects will be discussed elsewhere.

{\bf Acknowledgements}
S.T. would like to thank S.-K.Yang for encouragement.
The work of K.I. is supported in part by University of Tsukuba
Research Projects and the Grant-in-Aid for Scientific Research from
the Ministry of Education (No. 06740201 and No. 06221208).
The calculation of OPE has been done with the help of
Mathematica Package OPEdefs.m \cite{Th}.

{\bf Table 1} \quad The $W_{6}$ current of the  $WB_{3}$ algebra
{\tiny
$$
\begin{array}{|l|l||l|l|}
\hline \mbox{partition} & d^6(\mbox{partition}) & \mbox{partition}
 & d^6(\mbox{partition}) \\ \hline
(1^{6}    ,
     ,      ) & 1
 & (    ,1^{6}
    ,      ) & 1
 \\   (    ,
,1^{6}      ) & 1
 & (1^{4}    ,1^{2}
    ,      ) & -{{667}/ {235}}
 \\   (1^{4}    ,
,1^{2}      ) & -{{667}/ {235}}
 & (1^{2}    ,1^{4}
,      ) & -{{667}/ {235}}
 \\   (    ,1^{4}
,1^{2}      ) & -{{667}/ {235}}
 & (1^{2}    ,
,1^{4}      ) & -{{667}/ {235}}
 \\   (    ,1^{2}
,1^{4}      ) & -{{667}/ {235}}
 & (1^{2},\! 1^{2},\! 1^{2}) \! & {{5526}/ {235}}
 \\   (1^{4} 2   ,
,      ) &
{{3\,\left( 5 - 6\,{a^2} \right) }/ a}
 & ( 2   ,1^{4}
,      ) &
{{(-591 + 1258\,{a^2})}/ {235\,a}}
 \\   ( 2   ,
,1^{4}      ) &
 {{(-591 + 1258\,{a^2})}/ {235\,a}}
 & (1^{2} 2   ,1^{2}
,      ) &
{{2\,\left( -1963 + 2630\,{a^2} \right) }/ {235\,a}}
 \\   (1^{2} 2   ,
,1^{2}      ) &
{{2\,\left( -1963 + 2630\,{a^2} \right) }/ {235\,a}}
 & ( 2   ,1^{2}
,1^{2}      ) &
{{2\,\left( 95 - 2858\,{a^2} \right) }/ {235\,a}}
 \\   (1^{4}    , 2
,      ) &
{{667\,\left( -3 + 4\,{a^2} \right) }/ {235\,a}}
 & (    ,1^{4} 2
    ,      ) &
    {{3\,\left( 3 - 4\,{a^2} \right) }/ a}
 \\   (    , 2
    ,1^{4}      ) &
    {{(743 - 76\,{a^2})}/ {235\,a}}
 & (1^{3}    ,1 2
    ,      ) &
    {{2744\,\left( -1 + a^2 \right) \, }/ {235\,a}}
 \\   (1    ,1^{3} 2
    ,      ) &
    {{2744\,\left( -1 + a^2 \right) \, }/ {235\,a}}
 & (1^{2}    ,1^{2} 2
    ,      ) &
    {{1334\,\left( -3 + 4\,{a^2} \right) }/ {235\,a}}
 \\   (1^{2}    , 2
    ,1^{2}      ) &
    {{2\,\left( 1429 - 4192\,{a^2} \right) }/ {235\,a}}
 & (    ,1^{2} 2
    ,1^{2}      ) &
    {{2\,\left( -629 + 1296\,{a^2} \right) }/ {235\,a}}
 \\   (1^{4}    ,
    , 2     ) &
    {{667\,\left( -1 + 2\,{a^2} \right) }/ {235\,a}}
 & (    ,1^{4}
    , 2     ) &
    {{667\,\left( -1 + 2\,{a^2} \right) }/ {235\,a}}
 \\   (    ,
    ,1^{4} 2     ) &
    {{3\,\left( 1 - 2\,{a^2} \right) }/ a}
 & (1^{3}    ,
    ,1 2     ) &
    {{2744\,\left( -1 + a^2 \right) \, }/ {235\,a}}
 \\   (    ,1^{3}
    ,1 2     ) &
    {{2744\,\left( -1 + a^2 \right) \, }/ {235\,a}}
 & (1    ,
    ,1^{3} 2     ) &
    {{2744\,\left( -1 + a^2 \right) \, }/ {235\,a}}
 \\   (    ,1
    ,1^{3} 2     ) &
    {{2744\,\left( -1 + a^2 \right) \, }/ {235\,a}}
 & (1^{2}    ,1^{2}
    , 2     ) &
    {{5526\,\left( 1 - 2\,{a^2} \right) }/ {235\,a}}
 \\   (1^{2}    ,
    ,1^{2} 2     ) &
    {{1334\,\left( -1 + 2\,{a^2} \right) }/ {235\,a}}
 & (    ,1^{2}
    ,1^{2} 2     ) &
    {{1334\,\left( -1 + 2\,{a^2} \right) }/ {235\,a}}
 \\   (1^{2} 2^{2}   ,
    ,      ) &
    {{(115614 - 244181\,{a^2} + 114450\,{a^4})}/ {1645\,{a^2}}}
 & ( 2^{2}   ,1^{2}
    ,      ) &
    {{(-33386 + 45409\,{a^2} - 10696\,{a^4})}/ {1645\,{a^2}}}
 \\   ( 2^{2}   ,
    ,1^{2}      ) &
    {{(-33386 + 45409\,{a^2} - 10696\,{a^4})}/ {1645\,{a^2}}}
  & (1^{2}    , 2^{2}
    ,      ) &
    {{(-35514 + 40593\,{a^2} - 3752\,{a^4})}/ {1645\,{a^2}}}
  \\   (    ,1^{2} 2^{2}
    ,      ) &
    {{(55862 - 114541\,{a^2} + 44562\,{a^4})}/ {1645\,{a^2}}}
  & (    , 2^{2}
    ,1^{2}      ) &
    {({22110 - 36239\,{a^2} + 15456\,{a^4})}/ {1645\,{a^2}}}
  \\   (1    ,1 2^{2}
    ,      ) &
    {{2744\,\left( -1 + a^2 \right) \, \,
        \left( 3 - 4\,{a^2} \right) }/ {235\,{a^2}}}
  & (1^{2}    ,
    , 2^{2}     ) &
    {{(1838 - 4767\,{a^2} + 4256\,{a^4})}/ {1645\,{a^2}}}
   \\   (    ,1^{2}
    , 2^{2}     ) &
    {{(1838 - 4767\,{a^2} + 4256\,{a^4})}/ {1645\,{a^2}}}
  & (    ,
    ,1^{2} 2^{2}     ) &
    {{(16382 - 25445\,{a^2} - 5054\,{a^4})}/ {1645\,{a^2}}}
  \\   (1    ,1
    , 2^{2}     ) &
    {{16464\,{{\left( -1 + a^2 \right) }^2}\,{{ }^2}}/
      {235\,{a^2}}}
  & (1    ,
    ,1 2^{2}     ) &
    {{2744\,\left( -1 + a^2 \right) \, \,
        \left( 1 - 2\,{a^2} \right) }/ {235\,{a^2}}}
  \\   (    ,1
    ,1 2^{2}     ) &
    {{2744\,\left( -1 + a^2 \right) \, \,
        \left( 1 - 2\,{a^2} \right) }/ {235\,{a^2}}}
  & (1^{2} 2   , 2
    ,      ) &
    {{2\,\left( 1963 - 2630\,{a^2} \right) \,
        \left( -3 + 4\,{a^2} \right) }/ {235\,{a^2}}}
  \\   ( 2   ,1^{2} 2
    ,      ) &
    {{2\,\left( 591 - 1258\,{a^2} \right) \,
        \left( -3 + 4\,{a^2} \right) }/ {235\,{a^2}}}
  & ( 2   , 2
    ,1^{2}      ) &
    {{2\,\left( -1087 + 650\,{a^2} + 3200\,{a^4} \right) }/ {235\,{a^2}}}
  \\   (1 2   ,1 2
    ,      ) &
    {{4\,\left( -33888 + 61691\,{a^2} - 23863\,{a^4} \right) }/
      {1645\,{a^2}}}
  & (1^{2} 2   ,
    , 2     ) &
    {{2\,\left( 1963 - 2630\,{a^2} \right) \,
        \left( -1 + 2\,{a^2} \right) }/ {235\,{a^2}}}
  \\   ( 2   ,1^{2}
    , 2     ) &
    {{2\,\left( 95 - 2858\,{a^2} \right) \,\left( 1 - 2\,{a^2} \right) }
      / {235\,{a^2}}}
  & ( 2   ,
    ,1^{2} 2     ) &
    {{2\,\left( 591 - 1258\,{a^2} \right) \,
        \left( -1 + 2\,{a^2} \right) }/ {235\,{a^2}}}
  \\   (1 2   ,
    ,1 2     ) &
    {{4\,\left( -33888 + 61691\,{a^2} - 23863\,{a^4} \right) }/
      {1645\,{a^2}}}  &
( 2   ,1
    ,1 2     ) &
    {{2744\,\left( 1 - a \right) \, \,
        \left( 1 - 6\,{a^2} \right) }/ {235\,{a^2}}}
  \\   (1^{2}    , 2
    , 2     ) &
    {{2\,\left( 1429 - 4192\,{a^2} \right) \,
        \left( 1 - 2\,{a^2} \right) }/ {235\,{a^2}}}
  &  (    ,1^{2} 2
    , 2     ) &
    {{2\,\left( 629 - 1296\,{a^2} \right) \,
        \left( -1 + 2\,{a^2} \right) }/ {235\,{a^2}}}
  \\   (    , 2
    ,1^{2} 2     ) &
    {{2\,\left( -1 + 2\,{a^2} \right) \,
        \left( -743 + 76\,{a^2} \right) }/ {235\,{a^2}}}
   &  (1    ,1 2
    , 2     ) &
    {{2744\,\left( -1 + a^2 \right) \, \,
        \left( -1 + 2\,{a^2} \right) }/ {47\,{a^2}}}
  \\   (1    , 2
    ,1 2     ) &
    {{2744\,\left( -1 + a^2 \right) \, \,
        \left( -9 + 14\,{a^2} \right) }/ {235\,{a^2}}}
  &  (    ,1 2
    ,1 2     ) &
    {{4\,\left( 4528 - 15141\,{a^2} + 14553\,{a^4} \right) }/
      {1645\,{a^2}}}
  \\   ( 2^{3}   ,
    ,      ) &
    {{\left( -5 + 6\,{a^2} \right) \,
        \left( -17448 + 26103\,{a^2} - 5320\,{a^4} \right) }/
      {1645\,{a^3}}}
  &  (    , 2^{3}
    ,      ) &
    {{\left( -3 + 4\,{a^2} \right) \,
        \left( -10336 + 14903\,{a^2} - 1232\,{a^4} \right) }/
      {1645\,{a^3}}}
  \\   (    ,
    , 2^{3}     ) &
    {{\left( -1 + 2\,{a^2} \right) \,
        \left( 2824 - 8393\,{a^2} + 8904\,{a^4} \right) }/ {1645\,{a^3}}}
  &  ( 2^{2}   , 2
    ,      ) &
    {{\left( -3 + 4\,{a^2} \right) \,
        \left( 33386 - 45409\,{a^2} + 10696\,{a^4} \right) }/
      {1645\,{a^3}}}
  \\   ( 2^{2}   ,
    , 2     ) &
    {{\left( -1 + 2\,{a^2} \right) \,
        \left( 33386 - 45409\,{a^2} + 10696\,{a^4} \right) }/
      {1645\,{a^3}}}
  &  ( 2^{2}   ,
    , 2     ) &
    {{\left( -1 + 2\,{a^2} \right) \,
        \left( 33386 - 45409\,{a^2} + 10696\,{a^4} \right) }/
      {1645\,{a^3}}}
  \\   ( 2   , 2^{2}
    ,      ) &
    {{(-119946 + 358425\,{a^2} - 300734\,{a^4} + 60928\,{a^6})}/
      {1645\,{a^3}}}
   &  (    , 2^{2}
    , 2     ) &
    {{\left( -1 + 2\,{a^2} \right) \,
        \left( -22110 + 36239\,{a^2} - 15456\,{a^4} \right) }/
      {1645\,{a^3}}}
  \\   ( 2   ,
    , 2^{2}     ) &
    {{(-86850 + 214841\,{a^2} - 142198\,{a^4} + 12880\,{a^6})}/
      {1645\,{a^3}}}
  &  (    , 2
    , 2^{2}     ) &
    {{(24722 - 117693\,{a^2} + 185500\,{a^4} - 93856\,{a^6})}/
      {1645\,{a^3}}}
  \\   ( 2   , 2
    , 2     ) &
    {{2\,\left( 1 - 2\,{a^2} \right) \,
        \left( -1087 + 650\,{a^2} + 3200\,{a^4} \right) }/ {235\,{a^3}}}
  &  (1^{3}  3  ,
    ,      ) &
    {{(33947 - 71197\,{a^2} + 42140\,{a^4})}/ {1645\,{a^2}}}
  \\   (1  3  ,1^{2}    ,      ) &
    {{(-12701 + 44051\,{a^2} - 34692\,{a^4})}/ {1645\,{a^2}}}
  &  (1  3  ,    ,1^{2}      ) &
    {{(-12701 + 44051\,{a^2} - 34692\,{a^4})}/ {1645\,{a^2}}}
  \\   (1^{3}    ,  3  ,      ) &
    {{1372\,\left( -1 + a^2 \right) \, \,
        \left( 3 - 4\,{a^2} \right) }/ {235\,{a^2}}}
  &  (    ,1^{3}  3  ,      ) &
    {{(5135 - 3969\,{a^2} + 3724\,{a^4})}/ {1645\,{a^2}}}
  \\   (1^{2}    ,1  3  ,      ) &
    {{(-41513 + 111279\,{a^2} - 73108\,{a^4})}/ {1645\,{a^2}}}
  &  (1    ,1^{2}  3  ,      ) &
    {{1372\,\left( -1 + a^2 \right) \, \,
        \left( 3 - 4\,{a^2} \right) }/ {235\,{a^2}}}
   \\   (1    ,  3  ,1^{2}      ) &
    {{1372\,\left( 1 - a \right) \, \,
        \left( 3 - 8\,{a^2} \right) }/ {235\,{a^2}}}
  &  (    ,1  3  ,1^{2}      ) &
    {{(16111 - 61593\,{a^2} + 42140\,{a^4})}/ {1645\,{a^2}}}
  \\   (1^{3}    ,    ,  3    ) &
    {{1372\,\left( -1 + a^2 \right) \, \,
        \left( 1 - 2\,{a^2} \right) }/ {235\,{a^2}}}
  &  (    ,1^{3}    ,  3    ) &
    {{1372\,\left( -1 + a^2 \right) \, \,
        \left( 1 - 2\,{a^2} \right) }/ {235\,{a^2}}}
  \\   (    ,    ,1^{3}  3    ) &
    {{(-4469 + 24843\,{a^2} - 15484\,{a^4})}/ {1645\,{a^2}}}
  &  (1^{2}    ,1    ,  3    ) &
    {{1372\,\left( -1 + a^2 \right) \, \,
        \left( -1 + 2\,{a^2} \right) }/ {47\,{a^2}}}
  \\   (1^{2}    ,    ,1  3    ) &
    {{3\,\left( 2169 - 10927\,{a^2} + 7644\,{a^4} \right) }/
      {1645\,{a^2}}}
  &  (1    ,1^{2}    ,  3    ) &
    {{1372\,\left( -1 + a^2 \right) \, \,
        \left( -1 + 2\,{a^2} \right) }/ {47\,{a^2}}}
  \\   (    ,1^{2}    ,1  3    ) &
    {{3\,\left( 2169 - 10927\,{a^2} + 7644\,{a^4} \right) }/
      {1645\,{a^2}}}
  &  (1    ,    ,1^{2}  3    ) &
    {{1372\,\left( -1 + a^2 \right) \, \,
        \left( 1 - 2\,{a^2} \right) }/ {235\,{a^2}}}
   \\   (    ,1    ,1^{2}  3    ) &
    {{1372\,\left( -1 + a^2 \right) \, \,
        \left( 1 - 2\,{a^2} \right) }/ {235\,{a^2}}}
  &  (1 2 3  ,    ,      ) &
    {{\left( -5 + 6\,{a^2} \right) \,
        \left( -49863 + 91875\,{a^2} - 32830\,{a^4} \right) }/
      {1645\,{a^3}}}
  \\   (1  3  , 2   ,      ) &
    {{\left( -3 + 4\,{a^2} \right) \,
        \left( 12701 - 44051\,{a^2} + 34692\,{a^4} \right) }/
      {1645\,{a^3}}}
  &  (  3  ,1 2   ,      ) &
   {{2 \left( -37042 + 110473 {a^2} - 105301 {a^4} + 27930 {a^6} \right)
 }/ {1645 {a^3}}}
  \\   (  3  ,    ,1 2     ) &
    {{2\,\left( -37042 + 110473\,{a^2} - 105301\,{a^4} + 27930\,{a^6}
\right) }/ {1645\,{a^3}}}
  &  (1  3  ,    , 2     ) &
    {{\left( -1 + 2\,{a^2} \right) \,
        \left( 12701 - 44051\,{a^2} + 34692\,{a^4} \right) }/
      {1645\,{a^3}}}
  \\   (  3  ,    ,1 2     ) &
    {{2\,\left( -37042 + 110473\,{a^2} - 105301\,{a^4} + 27930\,{a^6}
\right) }/ {1645\,{a^3}}}
   &  (1 2   ,  3  ,      ) &
    {{2\,\left( -3 + 4\,{a^2} \right) \,
        \left( 33888 - 61691\,{a^2} + 23863\,{a^4} \right) }/
      {1645\,{a^3}}}
  \\   ( 2   ,1  3  ,      ) &
    {{(-92317 + 267649\,{a^2} - 264894\,{a^4} + 92904\,{a^6})}/
      {1645\,{a^3}}}
  &  (1    , 2 3  ,      ) &
    {{196\,\left( -1 + a^2 \right) \, \,
        \left( 81 - 112\,{a^2} + 28\,{a^4} \right) }/ {235\,{a^3}}}
  \\   (    ,1 2 3  ,      ) &
    {{\left( -3 + 4\,{a^2} \right) \,
        \left( -21051 + 24647\,{a^2} + 5586\,{a^4} \right) }/
      {1645\,{a^3}}}
  &  (    ,  3  ,1 2     ) &
    {{2\left( 40338 - 157517\,{a^2} + 190659\,{a^4} - 77420\,{a^6}
\right) }/ {1645\,{a^3}}} \\
(1    ,  3  , 2     ) &
    {{1372\,\left( -1 - a \right) \,\left( 1 - a \right) \,
        \left( 3 - 8\,{a^2} \right) \,\left( -1 + 2\,{a^2} \right) }/
      {235\,{a^3}}}
  &  (    ,1  3  , 2     ) &
    {{\left( -1 + 2\,{a^2} \right) \,
        \left( -16111 + 61593\,{a^2} - 42140\,{a^4} \right) }/
      {1645\,{a^3}}} \\ \hline
\end{array}
$$
$$
\begin{array}{|l|l||l|l|} \hline
     (    ,  3  ,1 2     ) &
    {{2\left( 40338 - 157517\,{a^2} + 190659\,{a^4} - 77420\,{a^6}
 \right) }/ {1645\,{a^3}}}
  &  (1 2   ,    ,  3    ) &
    {{2\,\left( -1 + 2\,{a^2} \right) \,
        \left( 33888 - 61691\,{a^2} + 23863\,{a^4} \right) }/
      {1645\,{a^3}}} \\
      ( 2   ,1    ,  3    ) &
    {{1372\,\left( -1 - a \right) \,\left( 1 - a \right) \,
        \left( 1 - 6\,{a^2} \right) \,\left( -1 + 2\,{a^2} \right) }/
      {235\,{a^3}}}
  &  ( 2   ,    ,1  3    ) &
    {{(-82713 + 181213\,{a^2} - 72814\,{a^4} - 22344\,{a^6})}/
      {1645\,{a^3}}}
  \\   (1    , 2   ,  3    ) &
    {{1372\,\left( -1 - a \right) \,\left( 1 - a \right) \,
        \left( 9 - 14\,{a^2} \right) \,\left( -1 + 2\,{a^2} \right) }/
      {235\,{a^3}}}
  &  (    ,1 2   ,  3    ) &
    {{2\left( -1 + 2 {a^2} \right) \,
        \left( -4528 + 15141\,{a^2} - 14553\,{a^4} \right) }/
      {1645\,{a^3}}}
  \\   (    ,    ,1 2 3    ) &
    {{\left( 1 - 2\,{a^2} \right) \,
        \left( 11447 + 4165\,{a^2} - 24794\,{a^4} \right) }/
      {1645\,{a^3}}}
  &  (1    ,    , 2 3    ) &
    {{196\,\left( -1 - a \right) \,\left( 1 - a \right) \,
        \left( -17 + 70\,{a^2} - 56\,{a^4} \right) }/ {235\,{a^3}}}
  \\   (    ,1    , 2 3    ) &
    {{196\,\left( -1 - a \right) \,\left( 1 - a \right) \,
        \left( -17 + 70\,{a^2} - 56\,{a^4} \right) }/ {235\,{a^3}}}
  &  (    ,    ,1 2 3    ) &
    {{\left( 1 - 2\,{a^2} \right) \,
        \left( 11447 + 4165\,{a^2} - 24794\,{a^4} \right) }/
      {1645\,{a^3}}}  \\ \hline
\end{array}
$$
$$
\begin{array}{|l|l|} \hline
 (  3^{2}  ,    ,      ) &
    {{(164846 - 664866\,{a^2} + 946483\,{a^4} - 551593\,{a^6} +
        106820\,{a^8}}/ {3290\,{a^4})}}  \\
(    ,  3^{2}  ,      ) &
    {{(26542 - 141610\,{a^2} + 217943\,{a^4} - 109025\,{a^6} +
        7840\,{a^8}}/ {3290\,{a^4})}}
   \\   (    ,    ,  3^{2}    ) &
    {{(-26146 + 31662\,{a^2} + 87363\,{a^4} - 154889\,{a^6} +
        63700\,{a^8})}/ {3290\,{a^4}}}
  \\   (  3  ,  3  ,      ) &
    {{\left( -3 + 4\,{a^2} \right) \,
        \left( 37042 - 110473\,{a^2} + 105301\,{a^4} - 27930\,{a^6} \right) }
       / {1645\,{a^4}}}
  \\   (  3  ,    ,  3    ) &
    {{\left( 1 - 2\,{a^2} \right) \,
        \left( -37042 + 110473\,{a^2} - 105301\,{a^4} + 27930\,{a^6} \right)
        }/ {1645\,{a^4}}}
  \\   (    ,  3  ,  3    ) &
    {{\left( 1 - 2\,{a^2} \right) \,
        \left( 40338 - 157517\,{a^2} + 190659\,{a^4} - 77420\,{a^6} \right) }
       / {1645\,{a^4}}}
  \\   (1^{2}   4 ,    ,      ) &
    {{\left( -5 + 6\,{a^2} \right) \,
        \left( -44217 + 69797\,{a^2} - 30380\,{a^4} \right) }/
      {9870\,{a^3}}}
  \\   (   4 ,1^{2}    ,      ) &
    {{(-25875 + 118361\,{a^2} - 140014\,{a^4} + 48216\,{a^6})}/
      {9870\,{a^3}}}
  \\   (   4 ,    ,1^{2}      ) &
    {{(-25875 + 118361\,{a^2} - 140014\,{a^4} + 48216\,{a^6})}/
      {9870\,{a^3}}}
   \\   ( 2  4 ,    ,      ) &
    {{(837297 - 3054953\,{a^2} + 3900286\,{a^4} - 1953798\,{a^6} +
        270480\,{a^8})}/ {9870\,{a^4}}}
  \\   (   4 , 2   ,      ) &
    {{\left( 3 - 4\,{a^2} \right) \,
        \left( -25875 + 118361\,{a^2} - 140014\,{a^4} + 48216\,{a^6} \right)
        }/ {9870\,{a^4}}}
  \\   (   4 ,    , 2     ) &
    {{\left( 1 - 2\,{a^2} \right) \,
        \left( -25875 + 118361\,{a^2} - 140014\,{a^4} + 48216\,{a^6} \right)
        }/ {9870\,{a^4}}}
  \\   (1^{2}    ,   4 ,      ) &
    {{\left( -3 + 4\,{a^2} \right) \,
        \left( 76519 - 151095\,{a^2} + 75264\,{a^4} \right) }/
      {9870\,{a^3}}}
  \\   (    ,1^{2}   4 ,      ) &
    {{\left( 3 - 4\,{a^2} \right) \,
        \left( 5801 + 16639\,{a^2} - 17640\,{a^4} \right) }/
      {9870\,{a^3}}}
  \\   (    ,   4 ,1^{2}      ) &
    {{(50331 - 192807\,{a^2} + 303100\,{a^4} - 159936\,{a^6})}/
      {9870\,{a^3}}}
  \\   (1    ,1   4 ,      ) &
    {{392\,\left( -1 + a^2 \right) \, \,
        \left( 51 - 133\,{a^2} + 84\,{a^4} \right) }/ {705\,{a^3}}}
  \\   ( 2   ,   4 ,      ) &
    {{\left( -3 + 4\,{a^2} \right) \,
        \left( 267347 - 715181\,{a^2} + 552986\,{a^4} - 105840\,{a^6}
\right) }/ {9870\,{a^4}}}
   \\   (    , 2  4 ,      ) &
    {{(-150063 + 683983\,{a^2} - 1117470\,{a^4} + 766318\,{a^6} -
        183456\,{a^8})}/ {9870\,{a^4}}}
  \\   (    ,   4 , 2     ) &
    {{\left( 1 - 2\,{a^2} \right) \,
        \left( 50331 - 192807\,{a^2} + 303100\,{a^4} - 159936\,{a^6} \right)
        }/ {9870\,{a^4}}}
  \\   (1^{2}    ,    ,   4   ) &
    {{\left( -1 + 2\,{a^2} \right) \,
        \left( -19521 + 89005\,{a^2} - 68796\,{a^4} \right) }/
      {9870\,{a^3}}}
  \\   (    ,1^{2}    ,   4   ) &
    {{\left( -1 + 2\,{a^2} \right) \,
        \left( -19521 + 89005\,{a^2} - 68796\,{a^4} \right) }/
      {9870\,{a^3}}}
  \\   (    ,    ,1^{2}   4   ) &
    {{\left( -1 + 2\,{a^2} \right) \,
        \left( 4469 - 21553\,{a^2} + 15484\,{a^4} \right) }/
      {3290\,{a^3}}}
  \\   (1    ,1    ,   4   ) &
    {{8232\,{{\left( -1 + a^2 \right) }^2}\,{{ }^2}\,
        \left( 1 - 2\,{a^2} \right) }/ {235\,{a^3}}}
  \\   (1    ,    ,1   4   ) &
    {{784\,\left( -1 + a^2 \right)  \,
        \left( -6 + 28\,{a^2} - 21\,{a^4} \right) }/ {705\,{a^3}}}
  \\   (    ,1    ,1   4   ) &
    {{784\,\left( -1 + a^2 \right) \,
        \left( -6 + 28\,{a^2} - 21\,{a^4} \right) }/ {705\,{a^3}}}
  \\   ( 2   ,    ,   4   ) &
    {{\left( -1 + 2\,{a^2} \right) \,
        \left( 248139 - 551913\,{a^2} + 236054\,{a^4} + 67032\,{a^6} \right)
        }/ {9870\,{a^4}}}
  \\   (    , 2   ,   4   ) &
    {{\left( 1 - 2\,{a^2} \right) \,
        \left( 58563 - 268267\,{a^2} + 485576\,{a^4} - 275184\,{a^6} \right)
        }/ {9870\,{a^4}}}
   \\   (    ,    , 2  4   ) &
    {{(-149823 + 429535\,{a^2} - 330394\,{a^4} - 93478\,{a^6} +
        143472\,{a^8})}/ {9870\,{a^4}}}
  \\   (1    5,    ,      ) &
    {{7\,\left( -1 + a^2 \right) \, \,
        \left( -612 + 1729\,{a^2} - 1414\,{a^4} + 280\,{a^6} \right) }/
      {235\,{a^4}}}
  \\   (1    ,    5,      ) &
    {{98\,\left( -1 + a^2 \right) \, \,
        \left( -3 + 4\,{a^2} \right) \,
        \left( -20 + 35\,{a^2} - 14\,{a^4} \right) }/ {235\,{a^4}}}
  \\   (    ,1    5,      ) &
    {{21\,\left( 1 - a^2 \right) \, \,
        \left( -76 + 287\,{a^2} - 378\,{a^4} + 168\,{a^6} \right) }/
      {235\,{a^4}}}
  \\   (1    ,    ,    5  ) &
    {{98\,\left( 1 - a^2 \right) \, \,
        \left( 1 - 2\,{a^2} \right) \,
        \left( 8 - 35\,{a^2} + 28\,{a^4} \right) }/ {235\,{a^4}}}
  \\   (    ,1    ,    5  ) &
    {{98\,\left( 1 - a^2 \right) \, \,
        \left( 1 - 2\,{a^2} \right) \,
        \left( 8 - 35\,{a^2} + 28\,{a^4} \right) }/ {235\,{a^4}}}
\\   (    ,    ,1    5  )&  {{7\,\left( -1 + a^2 \right) \,
        \left( 116 - 147\,{a^2} - 238\,{a^4} + 280\,{a^6} \right) }/
      {235\,{a^4}}}
   \\
 (    6,    ,      )&  {{\left( -5 + 6\,{a^2} \right) \,
        \left( -191688 + 633923\,{a^2} - 684691\,{a^4} + 263914\,{a^6} -
          27440\,{a^8} \right) }/ {98700\,{a^5}}}
  \\
 (    ,    6,      )& {{\left( -3 + 4\,{a^2} \right) \,
        \left( 280280 - 1119493\,{a^2} + 1614781\,{a^4} - 946190\,{a^6} +
          164640\,{a^8} \right) }/ {98700\,{a^5}}}
  \\
  (    ,    ,     6 )&  {{\left( 1 - 2\,{a^2} \right) \,
         \left( -170520 + 384101\,{a^2} + 86499\,{a^4} - 705698\,{a^6} +
                        411600\,{a^8} \right) }/ {98700\,{a^5}}}
  \hspace{1.93in} \\ \hline
\end{array}
$$
}
{\bf Table 2} \quad The $W_{6}$ current of $WG_{2}$ algebra
{\tiny
$$
\begin{array}{|l|l|} \hline \mbox{partition} & d^{6}(\mbox{partition})
 \hspace{4.95in} \\ \hline
 \left(1^{6}     ,      \right) &{{{a^6}\left( 2 - 9 {a^2} \right)
       \left( 1 - 2 {a^2} \right)
       \left( -6400 + 20103 {a^2} - 15552 {a^4} \right) }/{432}} \\
\left(      ,1^{6}     \right) &{{{a^6}\left( 2 - 3 {a^2} \right)
        \left( 3 - 2 {a^2} \right)
        \left( 1728 - 6701 {a^2} + 6400 {a^4} \right) }/{144}}
 \\  \left(1^{4}     ,1^{2}     \right)
&{{{a^6}\left( 118912 - 810842 {a^2} + 2038215 {a^4} -
           2238714 {a^6} + 905472 {a^8} \right) }/{144}}
  \\  \left(1^{2}     ,1^{4}    \right)
&{{{a^6}\left( -33536 + 248746 {a^2} - 679405 {a^4} +
           810842 {a^6} - 356736 {a^8} \right) }/{48}}
  \\  \left(1^{4} 2    ,     \right)
&{{\left( -1 - a \right) \left( 1 - a \right)  {a^5}
        \left( 2 - 9 {a^2} \right) \left( 1 - 2 {a^2} \right)
        \left( 6400 - 20103 {a^2} + 15552 {a^4} \right) }/
       {72 {\sqrt{2}}}}
  \\  \left(  2    ,1^{4}    \right)
&{{\left( -1 - a \right) \left( 1 - a \right)  {a^5}
        \left( 33536 - 248746 {a^2} + 679405 {a^4} - 810842 {a^6} +
           356736 {a^8} \right) }/{24 {\sqrt{2}}}}
  \\  \left(1^{3} 2    ,1 \right) &{{{a^5}\left( -2 + 3 {a^2} \right)
        \left( -112 + 387 {a^2} - 336 {a^4} \right)
        \left( 196 - 713 {a^2} + 588 {a^4} \right) }/{6 {\sqrt{6}}}}
  \\
\left(1  2    ,1^{3}\right) &{{{a^5}\left( 2 - 5 {a^2} \right)
        \left( -196 + 713 {a^2} - 588 {a^4} \right)
        \left( 112 - 387 {a^2} + 336 {a^4} \right) }/{6 {\sqrt{6}}}}
  \\ \hline
\end{array}
$$
$$
\begin{array}{|l|l|} \hline    \left(1^{2} 2    ,1^{2}\right)
&{{\left( -1 - a \right) \left( 1 - a \right)  {a^5}
        \left( -118912 + 810842 {a^2} - 2038215 {a^4} + 2238714 {a^6} -
           905472 {a^8} \right) }/{36 {\sqrt{2}}}}
  \\  \left(1^{4}     ,  2   \right)
&{{{a^5}\left( 67712 - 597154 {a^2} + 2099733 {a^4} -
           3651381 {a^6} + 3117690 {a^8} - 1036800 {a^{10}} \right) }/
       {72 {\sqrt{6}}}} \\
\left(      ,1^{4} 2   \right)
&{{{a^5}\left( 2 - 3 {a^2} \right)
        \left( 3 - 2 {a^2} \right) \left( -5 + 9 {a^2} \right)
        \left( -1728 + 6701 {a^2} - 6400 {a^4} \right) }/
       {24 {\sqrt{6}}}}
  \\  \left(1^{2}     ,1^{2} 2 \right)
&{{{a^5}\left( -79872 + 703202 {a^2} - 2448087 {a^4} +
           4223929 {a^6} - 3619746 {a^8} + 1234944 {a^{10}} \right) }/
       {12 {\sqrt{6}}}}
  \\  \left(1^{2} 2^{2}   ,     \right)
 &{{{a^4}\left( 131680 - 1338586 {a^2} + 5648195 {a^4} -
           12663269 {a^6} + 15903183 {a^8} - 10602090 {a^{10}} +
           2931552 {a^{12}} \right) }/{72}}
  \\  \left(  2^{2}   ,1^{2}    \right)
&{{{a^4}\left( -65216 + 1017734 {a^2} - 5958187 {a^4} +
           17418247 {a^6} - 27317505 {a^8} + 22012686 {a^{10}} -
           7162560 {a^{12}} \right) }/{72}}
  \\  \left(1  2^{2}   ,1 \right)
&{{\left( 1 - a \right)  {a^4}\left( 1 + a \right)
        \left( 2 - 3 {a^2} \right)
        \left( 112 - 387 {a^2} + 336 {a^4} \right)
        \left( 196 - 713 {a^2} + 588 {a^4} \right) }/{6 {\sqrt{3}}}}
  \\  \left(1^{2}     ,  2^{2}  \right)
&{{{a^4}\left( -56640 + 588586 {a^2} - 2533419 {a^4} +
           5784095 {a^6} - 7382049 {a^8} + 4992498 {a^{10}} -
           1401408 {a^{12}} \right) }/{72}}
  \\  \left(      ,1^{2} 2^{2}  \right) &{{{a^4}\left( 2 - 3 {a^2} \right)
        \left( 3 - 2 {a^2} \right)
        \left( 23376 - 170801 {a^2} + 467977 {a^4} - 571761 {a^6} +
           263376 {a^8} \right) }/{72}}
 \\  \left(1^{2} 2    ,  2 \right)
&{{\left( 1 - a \right)  {a^4}\left( 1 + a \right)
        \left( 67712 - 597154 {a^2} + 2099733 {a^4} - 3651381 {a^6} +
           3117690 {a^8} - 1036800 {a^{10}} \right) }/{36 {\sqrt{3}}}}
 \\  \left(  2    ,1^{2} 2 \right)
&{{\left( 1 - a \right)  {a^4}\left( 1 + a \right)
        \left( -79872 + 703202 {a^2} - 2448087 {a^4} + 4223929 {a^6} -
           3619746 {a^8} + 1234944 {a^{10}} \right) }/{12 {\sqrt{3}}}}
 \\  \left(1  2    ,1  2  \right)
&{{{a^4}\left( -33040 + 359780 {a^2} - 1620064 {a^4} +
           3867057 {a^6} - 5168870 {a^8} + 3672324 {a^{10}} -
           1083600 {a^{12}} \right) }/6}
 \\  \left(  2^{3}   ,     \right)
&{-{\left( 1 - a^{2} \right) {a^3}
        \left( -137280 + 1400402 {a^2} - 5946623 {a^4} + 13597517 {a^6} -
           17848893 {a^8} + 12848490 {a^{10}} - 3965760 {a^{12}} \right) }
        /{108 {\sqrt{2}}}}
 \\  \left(      ,  2^{3}    \right) &{{{a^3}\left( 3 - 2 {a^2} \right)
        \left( -2 + 3 {a^2} \right)
        \left( -19104 + 170473 {a^2} - 610976 {a^4} + 1101699 {a^6} -
           999225 {a^8} + 365472 {a^{10}} \right) }/{108 {\sqrt{6}}}}
 \\  \left(  2^{2}   ,  2 \right)
&{{{a^3}\left( -62656 + 909454 {a^2} - 5536793 {a^4} +
           18456374 {a^6} - 36502251 {a^8} + 42874803 {a^{10}} -
           27678078 {a^{12}} + 7563456 {a^{14}} \right) }/
       {36 {\sqrt{6}}}}
 \\  \left({  2    ,  2^{2}
     }\right) &{{-\left( 1 - a^{2} \right) {a^3}
        \left( 56640 - 588586 {a^2} + 2533419 {a^4} - 5784095 {a^6} +
           7382049 {a^8} - 4992498 {a^{10}} + 1401408 {a^{12}} \right) }
        /{36 {\sqrt{2}}}}
 \\  \left({1^{3}   3  ,
     }\right) &{{{a^4}\left( 75040 - 793904 {a^2} + 3492048 {a^4} -
           8161584 {a^6} + 10678311 {a^8} - 7406100 {a^{10}} +
           2122848 {a^{12}} \right) }/{108}}
 \\  \left({    3  ,1^{3}
     }\right) &{{\left( 1 - a^{2} \right)  {a^4}
        \left( 2 - 5 {a^2} \right)
        \left( -112 + 387 {a^2} - 336 {a^4} \right)
        \left( 196 - 713 {a^2} + 588 {a^4} \right) }/{12 {\sqrt{3}}}}
 \\  \left({1^{2}   3  ,1
     }\right) &{{\left( 1 - a^{2} \right)  {a^4}
        \left( 2 - 3 {a^2} \right)
        \left( 112 - 387 {a^2} + 336 {a^4} \right)
        \left( 196 - 713 {a^2} + 588 {a^4} \right) }/{12 {\sqrt{3}}}}
 \\  \left({1    3  ,1^{2}
     }\right) &{{{a^4}\left( -92064 + 1092656 {a^2} - 5273920 {a^4} +
           13291224 {a^6} - 18488667 {a^8} + 13483908 {a^{10}} -
           4034016 {a^{12}} \right) }/{36}}
 \\  \left({      ,1^{3}   3
     }\right) &{{{a^4}\left( 2 - 3 {a^2} \right)
        \left( 3 - 2 {a^2} \right)
        \left( 6608 - 50748 {a^2} + 145027 {a^4} - 182340 {a^6} +
           85008 {a^8} \right) }/{36}}
 \\  \left({1^{2}     ,1    3
     }\right) &
{{{a^4}\left( -48160 + 495864 {a^2} - 2105150 {a^4} +
           4721997 {a^6} - 5906274 {a^8} + 3908664 {a^{10}} -
           1070496 {a^{12}} \right) }/{36}}
 \\  \left({1  2  3  ,
     }\right) &
   {{-\left( 1 - a^{2} \right) {a^3}
        \left( -72240 + 750196 {a^2} - 3219428 {a^4} + 7286817 {a^6} -
           9141426 {a^8} + 6002964 {a^{10}} - 1605744 {a^{12}} \right) }
       /{18 {\sqrt{2}}}}
 \\  \left({  2  3  ,1
     }\right) &{{5 {a^5}\left( 2 - 3 {a^2} \right)
        \left( 1 - 2 {a^2} \right)
        \left( 112 - 387 {a^2} + 336 {a^4} \right)
        \left( 196 - 713 {a^2} + 588 {a^4} \right) }/{12 {\sqrt{6}}}}
 \\  \left({1    3  ,  2
     }\right) &{{{a^3}\left( -130368 + 1709744 {a^2} - 9495700 {a^4} +
           29004108 {a^6} - 52673817 {a^8} + 56916054 {a^{10}} -
           33906168 {a^{12}} + 8600256 {a^{14}} \right) }/
       {36 {\sqrt{6}}}}
 \\  \left({    3  ,1  2
     }\right) &{{-\left( 1 - a^{2} \right)  {a^3}
        \left( 33040 - 359780 {a^2} + 1620064 {a^4} - 3867057 {a^6} +
           5168870 {a^8} - 3672324 {a^{10}} + 1083600 {a^{12}} \right) }
       /{6 {\sqrt{2}}}}
 \\  \left({1  2    ,    3
     }\right) &
{{{a^3}\left( -113120 + 1388088 {a^2} - 7248616 {a^4} +
           20891364 {a^6} - 35940285 {a^8} + 36981270 {a^{10}} -
           21118752 {a^{12}} + 5171040 {a^{14}} \right) }/
       {36 {\sqrt{6}}}}
 \\  \left({  2    ,1    3
     }\right) &{{-\left( 1 - a^{2} \right)  {a^3}
        \left( 48160 - 495864 {a^2} + 2105150 {a^4} - 4721997 {a^6} +
           5906274 {a^8} - 3908664 {a^{10}} + 1070496 {a^{12}} \right) }
        /{18 {\sqrt{2}}}}
 \\  \left({      ,1  2  3
     }\right) &{{{a^3}\left( 2 - 3 {a^2} \right)
        \left( -3 + 2 {a^2} \right)
        \left( -10864 + 99324 {a^2} - 362902 {a^4} + 663211 {a^6} -
           607056 {a^8} + 222768 {a^{10}} \right) }/{12 {\sqrt{6}}}} \\
\left( {    3^{2}  ,
     }\right) &{a^2}\left( 95200 - 718280 {a^2} + 508692 {a^4} +
           12103458 {a^6} - 56831649 {a^8} \right. \\
 & \left. \hspace{.2in} + 121031121 {a^{10}} -
           140408586 {a^{12}} + 86215968 {a^{14}} - 22044960 {a^{16}}
           \right) /{432} \\
\left( {      ,    3^{2}
      }\right) &{{{a^2}\left( 2 - 3 {a^2} \right)
         \left( 3 - 2 {a^2} \right)
         \left( 20720 - 223548 {a^2} + 1004512 {a^4} - 2407563 {a^6} +
            3251043 {a^8} - 2349432 {a^{10}} + 710640 {a^{12}} \right) }
         /{432}} \\
\left( {    3  ,    3
      }\right) &\left(  1 - a^{2} \right)  {a^2}
         \left(  -113120 + 1388088 {a^2} - 7248616 {a^4} + 20891364 {a^6} -
            35940285 {a^8}  \right. \\
 & \left. \hspace{.5in}  + 36981270 {a^{10}} - 21118752 {a^{12}} +
            5171040 {a^{14}} \right) /{72 {\sqrt{3}}}
 \\  \left( {1^{2}    4 ,
      }\right) &{{-\left(  1 - a^{2} \right) {a^3}
         \left(  -150080 + 1575008 {a^2} - 6860690 {a^4} + 15915525 {a^6} -
            20792592 {a^8} + 14532264 {a^{10}} - 4245696 {a^{12}} \right) }
         /{216 {\sqrt{2}}}}
 \\
\left( {     4 ,1^{2}
     }\right) &{{-\left(  1 - a^{2} \right) {a^3}
        \left(  2 - 3 {a^2} \right)
        \left(  276192 - 2804224 {a^2} + 11210003 {a^4} - 22039560 {a^6} +
           21287304 {a^8} - 8068032 {a^{10}} \right) }/
       {216 {\sqrt{2}}}}
 \\  \left( {1     4 ,1
     }\right) &{{\left(  -1 - 2 a \right) \left(  1 - 2 a \right)  {a^3}
         {{\left(  -2 + 3 {a^2} \right) }^2}
        \left(  112 - 387 {a^2} + 336 {a^4} \right)
        \left(  196 - 713 {a^2} + 588 {a^4} \right) }/{36 {\sqrt{6}}}}
 \\  \left( {  2   4 ,
     }\right) &{a^2}\left(  24640 + 13264 {a^2} - 2263626 {a^4} +
           15398627 {a^6} - 49506115 {a^8} \right. \\
 & \left. \hspace{.2in} + 90542514 {a^{10}} -
           96858504 {a^{12}} + 56673864 {a^{14}} - 14043456 {a^{16}}
           \right) /{216}
 \\  \left( {     4 ,  2
     }\right) &\left(  1 - a^{2} \right)  {a^2}
        \left(  2 - 3 {a^2} \right)
        \left(  -195552 + 2237432 {a^2} - 10588825 {a^4} +
           26573058 {a^6} - 37325448 {a^8}  \right. \\
 & \left. \hspace{.5in} + 27827064 {a^{10}} -
           8600256 {a^{12}} \right) /{216 {\sqrt{3}}}
 \\  \left( {1^{2}     ,     4
     }\right) &{{{a^3}\left(  -1 + 2 {a^2} \right)
        \left(  42560 - 423104 {a^2} + 1733762 {a^4} - 3754133 {a^6} +
           4521420 {a^8} - 2863332 {a^{10}} + 743904 {a^{12}} \right) }
         /{72 {\sqrt{6}}}}
 \\  \left( {      ,1^{2}    4
     }\right) &{{{a^3}\left(  2 - 3 {a^2} \right)
        \left(  -3 + 2 {a^2} \right) \left(  -1 + 2 {a^2} \right)
        \left(  22176 - 169536 {a^2} + 479893 {a^4} - 593946 {a^6} +
           271152 {a^8} \right) }/{72 {\sqrt{6}}}}
 \\  \left( {  2    ,     4
     }\right) &{{\left(  1 - a^2 \right)  {a^2}
        \left(  1 - 2 {a^2} \right)
        \left(  -42560 + 423104 {a^2} - 1733762 {a^4} + 3754133 {a^6} -
           4521420 {a^8} + 2863332 {a^{10}} - 743904 {a^{12}} \right) }
        /{72 {\sqrt{3}}}}
 \\  \left( {      ,  2   4
     }\right) &{{{a^2}\left(  2 - 3 {a^2} \right)
        \left(  1 - 2 {a^2} \right) \left(  3 - 2 {a^2} \right)
        \left(  16800 - 151824 {a^2} + 548309 {a^4} - 990687 {a^6} +
           895194 {a^8} - 323568 {a^{10}} \right) }/{216}}
 \\  \left( {1      5,
     }\right) &{{{a^2}\left(  2 - 9 {a^2} \right)
        \left(  2 - 3 {a^2} \right) \left(  1 - 2 {a^2} \right)
        \left(  -3920 + 32256 {a^2} - 104968 {a^4} + 168849 {a^6} -
           134316 {a^8} + 42336 {a^{10}} \right) }/{108}}
 \\  \left( {      5,1
     }\right) &{{\left(  1 - a \right)  {a^2}\left(  1 + a \right)
        \left(  1 - 3 {a^2} \right) \left(  2 - 3 {a^2} \right)
        \left(  1 - 2 {a^2} \right)
        \left(  -112 + 387 {a^2} - 336 {a^4} \right)
        \left(  196 - 713 {a^2} + 588 {a^4} \right) }/{36 {\sqrt{3}}}}
 \\  \left(       ,1      5\right)
&{{{a^2}\left(  2 - 3 {a^2} \right)
        \left(  1 - 2 {a^2} \right) \left(  3 - 2 {a^2} \right)
        \left(  1568 - 14924 {a^2} + 56283 {a^4} - 104968 {a^6} +
           96768 {a^8} - 35280 {a^{10}} \right) }/{36}}
 \\  \left(       6,
     \right) &-\left(  1 - a^{2} \right) a
        \left(  2 - 3 {a^2} \right) \left(  1 - 2 {a^2} \right)
        \left(  117600 - 1405180 {a^2} + 6868151 {a^4}  \right.\\
 & \left. \hspace{1.4in} - 17605578 {a^6} +
           24993450 {a^8} - 18647496 {a^{10}} + 5715360 {a^{12}} \right)
        /{1620 {\sqrt{2}}}
 \\  \left(       ,
     6\right) &{{a\left(  2 - 3 {a^2} \right)
        \left(  1 - 2 {a^2} \right) \left(  3 - 2 {a^2} \right)
        \left(  -5 + 9 {a^2} \right)
        \left( -1568 + 14924 {a^2} - 56283 {a^4} + 104968 {a^6} -
           96768 {a^8} + 35280 {a^{10}} \right) }/{540 {\sqrt{6}}}} \\ \hline
\end{array}
$$
}

\end{document}